\documentclass[aps,jcp,reprint,superscriptaddress]{revtex4-1}
\usepackage{amsmath,amssymb}\usepackage{graphicx}\usepackage{subfigure}\usepackage{color}
\usepackage{times}\usepackage{tabularx}

\begin{document} 

\title{\bf {Note: A pairwise form of the Ewald sum for non-neutral systems}}
\author{Shasha Yi}
\affiliation{State Key Laboratory of Supramolecular Structure and Materials, College of Chemistry, Jilin University, Changchun, 130012, P. R. China}
\author{Cong Pan}
\affiliation{State Key Laboratory of Supramolecular Structure and Materials, College of Chemistry, Jilin University, Changchun, 130012, P. R. China}
\author{Zhonghan Hu} \email{zhonghanhu@jlu.edu.cn}
\affiliation{State Key Laboratory of Supramolecular Structure and Materials, College of Chemistry, Jilin University, Changchun, 130012, P. R. China}
\date{\today}

\begin{abstract}
Using an example of a mixed discrete-continuum representation of charges under the periodic boundary condition, we show that the exact pairwise form of the Ewald sum,
which is well-defined even if the system is non-neutral, provides a
natural starting point for deriving unambiguous Coulomb energies that must remove all spurious dependence on the choice of the Ewald screening factor.
\end{abstract}
\maketitle

In a recent article we derived a pairwise formulation for the Ewald sum associated with any inifinte boundary term\cite{Hu2014ib}. This formulation has an intuitive
interpretation of the contribution from the background charge that results in well-defined electrostatic energies. One of the main advantages of this formulation is
that, as opposed to other proposed derivations of the Ewald-type algorithm for non-neutral systems (e.g. \cite{Santos_Levin2016}), one can remove all spurious dependence of
the energy on the Ewald screening factor.

Let us consider a system of $N$ discrete point charges $(q_j,{\mathbf r}_j)$ with $j=1,2,\cdots,N$ in a cuboid unit cell specified by $L_x$, $L_y$, $L_z$ and their
infinite periodic images $(q_j,{\mathbf r}_j +{\mathbf n})$ where ${\mathbf n}$ stands for $(n_xL_x,n_yL_y,n_zL_z)$ with $n_x$, $n_y$, and $n_z$ integers. The usual
Ewald3D sum under the tinfoil boundary condition (e3dtf) for the Coulomb energy of the unit cell
reads\cite{Ewald1921,DeLeeuw_Smith1980,Frenkel_Smit2002}
\begin{multline} {\cal U}^{\rm e3dtf} = \frac{1}{2}\sum_{i,j=1}^N q_iq_j\sideset{}{'}\sum_{\mathbf n} \frac{{\rm erfc}(\alpha|{\mathbf r}_{ij}
+{\mathbf n}|)}{|{\mathbf r}_{ij}+{\mathbf n}|} \\
- \frac{\alpha}{\sqrt\pi} \sum_{j=1}^N q_j^2  + \frac{2\pi}{V}\sum_{{\mathbf k}\neq{\mathbf 0}} \frac{e^{-k^2/(4\alpha^2)}}{k^2}
\left|\sum_{j=1}^N q_j e^{i{\mathbf k}\cdot{\mathbf r}_j}\right|^2  , \end{multline}
where ${\mathbf r}_{ij}={\mathbf r}_i - {\mathbf r}_j$, $V=L_xL_yL_z$ and 
${\mathbf k}=2\pi (k_x/L_x,k_y/L_y, k_z/L_z)$ with $k_x$, $k_y$, and $k_z$ integers. The prime indicates that the $i=j$ terms are omitted when ${\mathbf
n}=0$. The parameter $\alpha\in(0,\infty)$ is a screening factor that determines the relative proportion of the real and reciprocal space sums. However, ${\cal
U}^{\rm e3dtf}$ uniformly and absolutely converges to an $\alpha$-independent value for any given non-overlapping configuration. Under the electroneutrality
condition, $\sum_{j=1}^N q_j = 0 $, 
${\cal U}^{\rm e3dtf}$ can be exactly re-expressed as a conventional pairwise form (see Fig. 1 and eqs. (28)-(34) of ref.\cite{Hu2014ib})
\begin{equation} {\cal U}^{\rm e3dtf} = \sum_{i<j}^N q_i q_j \nu^{\rm e3dtf}({\mathbf r}_i-{\mathbf r}_j) , \label{eq:pwU} \end{equation}
where
\begin{multline} \nu^{\rm e3dtf}({\mathbf r}) = \tau^{\rm 3D} + \sum_{\mathbf n}
 \frac{ {\rm erfc}(\alpha|{\mathbf r}+ {\mathbf n}|) }{\left|{\mathbf r} + {\mathbf n}\right|} +  \\ \frac{4\pi}{V} \sum_{{\mathbf k}\neq {\mathbf 0}}
\frac{e^{-k^2/(4\alpha^2)}e^{i{\mathbf k}\cdot{\mathbf r} }}{k^2} - \frac{\pi}{\alpha^2 V} . \label{eq:nu} \end{multline}
The constant $\tau^{\rm 3D}$  independent of ${\mathbf r}$ and $\alpha$ is given by
\begin{multline}  \tau^{\rm 3D}= \frac{\pi}{\alpha^2 V} + \frac{2\alpha}{\sqrt{\pi}} - \sum_{ {\mathbf n}\neq {\mathbf 0}} 
\frac{ {\rm erfc}(\alpha|{\mathbf n}|)}{|{\mathbf n}|} \\ -  \frac{4\pi}{V}\sum_{ {\mathbf k}\neq {\mathbf 0}} \frac{e^{-k^2/(4\alpha^2)} }{k^2}  \end{multline}
Both $\tau^{\rm 3D}$ and $\nu^{\rm e3dtf}({\mathbf r})$ absolutely and uniformly converge to $\alpha$-independent values for any $\alpha\in(0,\infty)$.
Taking $\alpha\to\infty$ in eq.~\eqref{eq:nu}, a more concise form for $\nu^{\rm e3dtf}({\mathbf r})$ formally reads
\begin{equation}
 \nu^{\rm e3dtf}({\mathbf r}) =  \tau^{\rm 3D} + \frac{4\pi}{V} \sum_{{\mathbf k}\neq{\mathbf 0}} \frac{e^{i{\mathbf k}\cdot{\mathbf r}}}{k^2} \label{eq:pwnu}.
 \end{equation}
When the system is non-neutral, $\sum_{j=1}^N q_j \neq 0$, ${\cal U}^{\rm e3dtf}$ of eq.(1) still converges but its value depends on $\alpha$. 
In contrast, the pairwise expression eq.~\eqref{eq:pwU} remains well-defined and independent of $\alpha$ because the effect of the background charges has been taken into account by 
$\nu^{\rm e3dtf}$. As will be shown below, $\nu^{\rm e3dtf}$ offers additional convenience when deriving any Ewald sum formula for a continuous distribution of
charges.


Rigorous derivations of the Ewald sum\cite{Smith1981,Smith2008,Ballenegger2014} have shown that the Coulomb energy of the unit cell inside an infinite periodic
lattice has an extra shape-dependent term that depends on the asymptotic behavior that the lattice approaches the infinite.
Alternatively, this infinite boundary term can be obtained transparently by an analysis of ${\mathbf k}\to0$ behavior of the reciprocal space term\cite{Hu2014ib}
\begin{equation} 
{\cal U}^{\rm e3d} = {\cal U}^{\rm e3dtf} - \frac{\pi}{V} \sum_{i,j}^N  q_iq_j \lim_{{\mathbf k}\to{\mathbf 0}} \frac{\left({\mathbf k}
\cdot{\mathbf r_{ij}}\right)^2}{k^2},
 \end{equation}
For example, regarding $\lim_{ {\mathbf k}\to 0}$ as $\lim_{k_z \to 0}\left[\lim_{k_x,k_y\to 0} \right]$ yields the Ewald sum with the planar
infinite boundary term\cite{Hu2014ib}
\begin{equation} {\cal U}^{\rm e3d}_{\rm p} = {\cal U}^{\rm e3dtf} - \frac{\pi}{V} \sum_{i,j}^N q_i q_jz_{ij}^2 = \sum_{i<j}^N q_iq_j
 \nu^{\rm e3d}_{\rm p}({\mathbf r}_{ij})  \end{equation}
where the corresponding pairwise potential is given by
\begin{equation} \nu^{\rm e3d}_{\rm p}({\mathbf r}) = \nu^{\rm e3dtf}({\mathbf r}) - \frac{2\pi}{V}z^2 . \label{eq:pwnup} \end{equation}
This planar infinite boundary term was actually known since 1980s\cite{Smith1981} and was later widely used as a correction to the usual Ewald3D sum with the tinfoil
boundary term when simulating planar interfaces \cite{Neugebauer_Scheffler1992,Yeh_Berkowitz1999}. Other efficient and accurate methods using mean-field ideas or the
2D periodic Ewald sum for such systems have been recently developed\cite{Hu_Weeks2010lmf,Lindbo_Tornberg2012,Hu2014spmf,Pan_Hu2014,Pan_Hu2015}.
Relations among them have been discussed\cite{Yi_Hu2015,Yi_Hu2017}. Moreover, it has been clarified that ${\cal U}^{\rm e3d}_{\rm p}$ is in fact an accurate
mean-field approximation to the 2D periodic Ewald sum\cite{Pan_Hu2017}.

In a recent example studied by Levin and coworkers\cite{Santos_Levin2016}, an efficient algorithm was developed to simulate a system of $N_m$ mobile ions
$(q_j,{\mathbf r}_j)$ confined between two charged planar walls by treating the $N-N_m$ fixed charges, $(q_s,{\mathbf r}_s)$ on the wall as continuum to
reduce the computational cost. Neither the fixed charges nor the mobile ions necessarily satisfy the electroneutrality condition.
We now use the above pairwise potential $\nu^{\rm e3d}_{\rm p}({\mathbf r})$ of eq.\eqref{eq:pwnup} as a starting point to simply and transparently derive the Ewald
sum energy that removes a spurious dependence on the screen factor $\alpha$ which appeared in Ref.\cite{Santos_Levin2016}.

Using the Ewald sum with the planar infinite boundary term, the total coulomb energy of the mobile ions can be written as a sum of the mobile-mobile and mobile-fixed
components
\begin{equation} {\cal U}^{\rm e3d}_{\rm p} = \sum_{i<j}^{N_m} q_iq_j\nu^{\rm e3d}_{\rm p}({\mathbf r}_{ij}) +  \sum_{j=1}^{N_m} q_j \phi_{\rm mf}({\mathbf r}_j),
 \end{equation}
where each mobile charge $(q_j,{\mathbf r}_j)$ interacts with the wall through the potential
\begin{equation} \phi_{\rm mf}({\mathbf r}) = \sum_{s=N_m+1}^N q_s \nu^{\rm e3d}_{\rm p}({\mathbf r} - {\mathbf r}_s). \end{equation}

As suggested by Levin and coworkers, the discrete fixed charges on the wall $(q_s,{\mathbf r}_s)$ might be replaced by a surface charge density distribution 
$\rho^q_s({\mathbf r})=\sigma_1\delta(z-z_L) + \sigma_2\delta(z-z_R)$. Consequently, one then approximates $\phi_{\rm mf}({\mathbf r}) $ as 
\begin{equation} \phi_{\rm mf}({\mathbf r}) \simeq \int_V d{\mathbf r}^\prime\, \nu^{\rm e3d}_{\rm p}({\mathbf r} - {\mathbf r}^\prime) 
\rho^q_s({\mathbf r}^\prime) = 2\pi  \left(\sigma_2-\sigma_1\right) z + C_1  \label{eq:ep}\end{equation}
where $C_1$, a constant independent of ${\mathbf r}$ is given by
\begin{equation} C_1 = (\sigma_1+\sigma_2)(L_xL_y\tau^{\rm 3D}+\frac{2\pi L_z}{6})+2\pi(\sigma_1 z_L - \sigma_2 z_R). \end{equation}
When integrating $\nu^{\rm e3d}_{\rm p}$ of eq.~\eqref{eq:pwnu} and~\eqref{eq:pwnup}, terms in eq.~\eqref{eq:pwnu} with $k_x\neq 0$ or
$k_y\neq 0$ all vanishes. The surviving terms with $k_x=k_y=0; k_z\neq 0$ are related to the Fourier series for $|z|-z^2/L_z$ on an interval $[-L_z,L_z]$
\begin{equation}  |z| - \frac{z^2}{L_z} = \frac{L_z}{6} - \frac{2}{L_z}\sum_{k_z\neq0} \frac{e^{i2\pi k_zz/L_z}}{(2\pi k_z/L_z)^2}.  \end{equation}

On the other hand, the mobile-mobile component can be rewritten in the usual form
\begin{multline} 
\sum_{i<j}^{N_m} q_iq_j\nu^{\rm e3d}_{\rm p}({\mathbf r}_{ij})= \frac{1}{2} \sum_{i\neq j}^{N_m} q_iq_j \sum_{\mathbf n} 
\frac{{\rm erfc}(\alpha|{\mathbf r}_{ij} +{\mathbf n}|)}{|{\mathbf r}_{ij}+{\mathbf n}|}  \\ 
+ \frac{2\pi}{V}\sum_{{\mathbf k}\neq{\mathbf 0}} \frac{e^{-k^2/(4\alpha^2)}}{k^2} \left|\sum_{j=1}^N q_j e^{i{\mathbf k}\cdot{\mathbf r}_j}\right|^2 + \\
+ \frac{2\pi}{V}\left[M_z^2 - Q_tG_z \right] + C_2(\alpha) \label{eq:e}
 \end{multline}
where $M_z$, $Q_t$ and $G_z$ are defined as the same in the ref.\cite{Santos_Levin2016}
\[ M_z = \sum_{j=1}^{N_m} q_j z_j ;\quad Q_t = \sum_{j=1}^{N_m} q_j; \quad G_z = \sum_{j=1}^{N_m} q_j z_j^2,  \]
and the $\alpha$-dependent constant is given by
\begin{equation} C_2(\alpha) = \sum_{i<j}^{N_m} q_iq_j\left( \tau^{\rm 3D} - \frac{\pi}{\alpha^2 V}  \right) - \frac{2\pi}{V}\sum_{j=1}^{N_m} q_j^2 \sum_{\mathbf k\neq 0}
\frac{e^{-k^2/(4\alpha^2)}}{k^2}. \end{equation}
Clearly, our expressions~\eqref{eq:e} and~\eqref{eq:ep} differ from the corresponding eqs.(19) and (21) of Ref.\cite{Santos_Levin2016} by constants $C_1$ and $C_2$
respectively. Both constants are useful for validating the approximations to the Coulomb energies. However, it should be noted that molecular dynamics or Monte-Carlo
simulation results\cite{Santos_Levin2016,Colla_Levin2016,Santos_Levin2016jpcb} should not depend on these constants because they are cancelled in the forces and relative energies.

We would like to thank Prof. Claudio Margulis for bringing to our attention ref.\cite{Santos_Levin2016}. This work was supported by the NSFC (grant no. 21522304) and the Program for JLUSTIRT.



\bibliographystyle{apsrev}
\end{document}